\documentclass[epj,nopacs]{svjoura}
\usepackage{amsmath}
\usepackage{graphics}
\usepackage{graphicx}
\usepackage{epsfig}
\usepackage{amssymb}
\usepackage{units}
\usepackage{cite}
\usepackage{lineno}
\usepackage{calc}
\usepackage{endnotes}
\usepackage{fixltx2e}
\usepackage{float}
\usepackage{multirow}
\usepackage{hyperref}
\usepackage{grffile}

\newcommand{\dNdEta}{{\rm d}N_{\rm ch}/{\rm d}\eta }
\newcommand{\bfig}[1][t]{\begin{figure}[#1] \begin{center}}
\newcommand{\efig}{\end{center} \end{figure}}

\newcommand{\btab}[1][ht!]{\begin{table}[#1] \begin{center}}
\newcommand{\etab}{\end{center} \end{table}}

\newcommand{\bq}{\begin{equation}}
\newcommand{\eq}{\end{equation}}

\newcommand{\etain}[1]{$|\eta|<#1$}
\newcommand{\cms}{\sqrt{s}}
\newcounter{vers}

\usepackage{preprintcover}
\PreprintIdNumber{}
\PreprintCoverPaperTitle{Charged-particle multiplicity measurement in proton-proton
collisions at ${\sqrt{{\it s}}= 7}$\,TeV with ALICE at LHC}
\PreprintCoverAbstract{The pseudorapidity density and multiplicity distribution of charged particles produced in proton--proton collisions at the LHC, at a centre-of-mass energy $\cms = 7$~TeV, were measured in the central pseudorapidity region \etain{1}. Comparisons are made with previous measurements at $\cms = 0.9$~TeV and 2.36~TeV. At $\cms = 7$~TeV, for events with at least one charged particle in \etain{1}, we obtain $\dNdEta = 6.01 \pm 0.01 (\emph{stat.}) ^{+0.20}_{-0.12} (\emph{syst.})$. This corresponds to an increase of $57.6\,\% \pm 0.4\,\%(\emph{stat.}) ^{+3.6}_{-1.8}\,\%(\emph{syst.})$ relative to collisions at 0.9~TeV, significantly higher than calculations from commonly used models. The multiplicity distribution at 7~TeV is described fairly well by the negative binomial distribution.
}
\PreprintJournalName{EPJC}

\begin{document}
\hugehead
\title{Charged-particle multiplicity measurement in proton-proton
collisions at $\mathbf{\sqrt{{\it s}}= 7}$\,TeV with ALICE at LHC}

\subtitle{ALICE collaboration}

\author{  
K.~Aamodt\inst{79} \and  
N.~Abel\inst{43} \and  
U.~Abeysekara\inst{77} \and  
A.~Abrahantes~Quintana\inst{42} \and  
A.~Abramyan\inst{113} \and  
D.~Adamov\'{a}\inst{87} \and  
M.M.~Aggarwal\inst{25} \and  
G.~Aglieri~Rinella\inst{40} \and  
A.G.~Agocs\inst{18} \and  
S.~Aguilar~Salazar\inst{65}
% ...
\renewcommand{\notesname}{Affiliation notes}  
\endnotetext[1]{Also at{ Laboratoire de Physique Corpusculaire (LPC), Clermont Universit\'{e}, Universit\'{e} Blaise Pascal, CNRS--IN2P3, Clermont-Ferrand, France}}  
\endnotetext[2]{Also at{ Frankfurt Institute for Advanced Studies, Johann Wolfgang Goethe-Universit\"{a}t Frankfurt, Frankfurt, Germany}}  
\endnotetext[3]{Now at{ Sezione INFN, Padova, Italy}}  
\endnotetext[4]{Now at{ Research Division and ExtreMe Matter Institute EMMI, GSI Helmholtzzentrum f\"{u}r Schwerionenforschung, Darmstadt, Germany}}  
\endnotetext[5]{Now at{ Institut f\"{u}r Kernphysik, Johann Wolfgang Goethe-Universit\"{a}t Frankfurt, Frankfurt, Germany}}  
\endnotetext[6]{Now at{ Physics Department, University of Cape Town, iThemba Laboratories, Cape Town, South Africa}}  
\endnotetext[7]{Now at{ National Institute for Physics and Nuclear Engineering, Bucharest, Romania}}  
\endnotetext[8]{Also at{ University of Houston, Houston, TX, United States}}  
\endnotetext[9]{Now at{ Faculty of Science, P.J.~\v{S}af\'{a}rik University, Ko\v{s}ice, Slovakia}}  
\endnotetext[10]{Now at{ European Organization for Nuclear Research (CERN), Geneva, Switzerland}}  
\endnotetext[11]{Now at{ Helsinki Institute of Physics (HIP) and University of Jyv\"{a}skyl\"{a}, Jyv\"{a}skyl\"{a}, Finland}}  
\endnotetext[12]{Now at{ Institut Pluridisciplinaire Hubert Curien (IPHC), Universit\'{e} de Strasbourg, CNRS-IN2P3, Strasbourg, France}}  
\endnotetext[13]{Now at{ Sezione INFN, Bari, Italy}}  
\endnotetext[14]{Now at{ Institut f\"{u}r Kernphysik, Westf\"{a}lische Wilhelms-Universit\"{a}t M\"{u}nster, M\"{u}nster, Germany}}  
\endnotetext[15]{Now at: University of Technology and Austrian Academy of Sciences, Vienna, Austria}  
\endnotetext[16]{Also at{ Lawrence Livermore National Laboratory, Livermore, CA, United States}}  
\endnotetext[17]{Also at{ European Organization for Nuclear Research (CERN), Geneva, Switzerland}}  
\endnotetext[18]{Now at { Secci\'{o}n F\'{\i}sica, Departamento de Ciencias, Pontificia Universidad Cat\'{o}lica del Per\'{u}, Lima, Peru}}  
\endnotetext[19]{Deceased}  
\endnotetext[20]{Now at{ Yale University, New Haven, CT, United States}}  
\endnotetext[21]{Now at{ University of Tsukuba, Tsukuba, Japan}}  
\endnotetext[22]{Also at { Centro Fermi -- Centro Studi e Ricerche e Museo Storico della Fisica ``Enrico Fermi'', Rome, Italy}}  
\endnotetext[23]{Now at{ Dipartimento Interateneo di Fisica `M.~Merlin' and Sezione INFN, Bari, Italy}}  
\endnotetext[24]{Also at{ Laboratoire de Physique Subatomique et de Cosmologie (LPSC), Universit\'{e} Joseph Fourier, CNRS-IN2P3, Institut Polytechnique de Grenoble, Grenoble, France}}  
\endnotetext[25]{Now at{ Dipartimento di Fisica Sperimentale dell'Universit\`{a} and Sezione INFN, Turin, Italy}}  
\endnotetext[26]{Now at{ Physics Department, Creighton University, Omaha, NE, United States}}  
\endnotetext[27]{Now at{ Commissariat \`{a} l'Energie Atomique, IRFU, Saclay, France}}  
\endnotetext[28]{Also at{ Department of Physics, University of Oslo, Oslo, Norway}}  
\endnotetext[29]{Now at{ Physikalisches Institut, Ruprecht-Karls-Universit\"{a}t Heidelberg, Heidelberg, Germany}}  
\endnotetext[30]{Now at{ Institut f\"{u}r Kernphysik, Technische Universit\"{a}t Darmstadt, Darmstadt, Germany}}  
\endnotetext[31]{Now at{ Department of Physics and Technology, University of Bergen, Bergen, Norway}}  
\endnotetext[32]{Now at{ Physics Department, University of Athens, Athens, Greece}}  
\endnotetext[33]{Also at{ Institut f\"{u}r Kernphysik, Westf\"{a}lische Wilhelms-Universit\"{a}t M\"{u}nster, M\"{u}nster, Germany}}  
\endnotetext[34]{Now at{ SUBATECH, Ecole des Mines de Nantes, Universit\'{e} de Nantes, CNRS-IN2P3, Nantes, France}}  
\endnotetext[35]{Now at{ Universit\'{e} de Lyon 1, CNRS/IN2P3, Institut de Physique Nucl\'{e}aire de Lyon, Lyon, France}}  
\endnotetext[36]{Now at: Centre de Calcul IN2P3, Lyon, France}  
\endnotetext[37]{Now at{ Variable Energy Cyclotron Centre, Kolkata, India}}  
\endnotetext[38]{Also at{ Dipartimento di Fisica dell'Universit\`{a} and Sezione INFN, Padova, Italy}}  
\endnotetext[39]{Also at{ Sezione INFN, Bologna, Italy}}  
\endnotetext[40]{Also at Dipartimento di Fisica dell\'{ }Universit\`{a}, Udine, Italy}  
\endnotetext[41]{Also at Wroc{\l}aw University, Wroc{\l}aw, Poland} 
\bigskip  
\theendnotes  
\section*{Collaboration institutes}  
}  

\institute{
Dipartimento di Scienze e Tecnologie Avanzate dell'Universit\`{a} del Piemonte Orientale and Gruppo Collegato INFN, Alessandria, Italy
\and
Department of Physics Aligarh Muslim University, Aligarh, India
\and
National Institute for Nuclear and High Energy Physics (NIKHEF), Amsterdam, Netherlands
\and
Physics Department, University of Athens, Athens, Greece
\and
Dipartimento Interateneo di Fisica `M.~Merlin' and Sezione INFN, Bari, Italy
\and
Sezione INFN, Bari, Italy
\and
China Institute of Atomic Energy, Beijing, China
\and
Department of Physics and Technology, University of Bergen, Bergen, Norway
\and
Faculty of Engineering, Bergen University College, Bergen, Norway
\and
Lawrence Berkeley National Laboratory, Berkeley, CA, United States
\and
Institute of Physics, Bhubaneswar, India
\and
School of Physics and Astronomy, University of Birmingham, Birmingham, United Kingdom
\and
Dipartimento di Fisica dell'Universit\`{a} and Sezione INFN, Bologna, Italy
\and
Sezione INFN, Bologna, Italy
\and
Faculty of Mathematics, Physics and Informatics, Comenius University, Bratislava, Slovakia
\and
Institute of Space Sciences (ISS), Bucharest, Romania
\and
National Institute for Physics and Nuclear Engineering, Bucharest, Romania
\and
KFKI Research Institute for Particle and Nuclear Physics, Hungarian Academy of Sciences, Budapest, Hungary
\and
Dipartimento di Fisica dell'Universit\`{a} and Sezione INFN, Cagliari, Italy
\and
Sezione INFN, Cagliari, Italy
\and
Universidade Estadual de Campinas (UNICAMP), Campinas, Brazil
\and
Physics Department, University of Cape Town, iThemba Laboratories, Cape Town, South Africa
\and
Dipartimento di Fisica e Astronomia dell'Universit\`{a} and Sezione INFN, Catania, Italy
\and
Sezione INFN, Catania, Italy
\and
Physics Department, Panjab University, Chandigarh, India
\and
Laboratoire de Physique Corpusculaire (LPC), Clermont Universit\'{e}, Universit\'{e} Blaise Pascal, CNRS--IN2P3, Clermont-Ferrand, France
\and
Department of Physics, Ohio State University, Columbus, OH, United States
\and
Niels Bohr Institute, University of Copenhagen, Copenhagen, Denmark
\and
The Henryk Niewodniczanski Institute of Nuclear Physics, Polish Academy of Sciences, Cracow, Poland
\and
Universidad Aut\'{o}noma de Sinaloa, Culiac\'{a}n, Mexico
\and
Research Division and ExtreMe Matter Institute EMMI, GSI Helmholtzzentrum f\"{u}r Schwerionenforschung, Darmstadt, Germany
\and
Institut f\"{u}r Kernphysik, Technische Universit\"{a}t Darmstadt, Darmstadt, Germany
\and
Wayne State University, Detroit, MI, United States
\and
Joint Institute for Nuclear Research (JINR), Dubna, Russia
\and
Frankfurt Institute for Advanced Studies, Johann Wolfgang Goethe-Universit\"{a}t Frankfurt, Frankfurt, Germany
\and
Institut f\"{u}r Kernphysik, Johann Wolfgang Goethe-Universit\"{a}t Frankfurt, Frankfurt, Germany
\and
Laboratori Nazionali di Frascati, INFN, Frascati, Italy
\and
Gangneung-Wonju National University, Gangneung, South Korea
\and
Petersburg Nuclear Physics Institute, Gatchina, Russia
\and
European Organization for Nuclear Research (CERN), Geneva, Switzerland
\and
Laboratoire de Physique Subatomique et de Cosmologie (LPSC), Universit\'{e} Joseph Fourier, CNRS-IN2P3, Institut Polytechnique de Grenoble, Grenoble, France
\and
Centro de Aplicaciones Tecnol\'{o}gicas y Desarrollo Nuclear (CEADEN), Havana, Cuba
\and
Kirchhoff-Institut f\"{u}r Physik, Ruprecht-Karls-Universit\"{a}t Heidelberg, Heidelberg, Germany
\and
Physikalisches Institut, Ruprecht-Karls-Universit\"{a}t Heidelberg, Heidelberg, Germany
\and
Hiroshima University, Hiroshima, Japan
\and
University of Houston, Houston, TX, United States
\and
Physics Department, University of Rajasthan, Jaipur, India
\and
Physics Department, University of Jammu, Jammu, India
\and
Helsinki Institute of Physics (HIP) and University of Jyv\"{a}skyl\"{a}, Jyv\"{a}skyl\"{a}, Finland
\and
Scientific Research Technological Institute of Instrument Engineering, Kharkov, Ukraine
\and
Bogolyubov Institute for Theoretical Physics, Kiev, Ukraine
\and
University of Tennessee, Knoxville, TN, United States
\and
Saha Institute of Nuclear Physics, Kolkata, India
\and
Variable Energy Cyclotron Centre, Kolkata, India
\and
Fachhochschule K\"{o}ln, K\"{o}ln, Germany
\and
Faculty of Science, P.J.~\v{S}af\'{a}rik University, Ko\v{s}ice, Slovakia
\and
Institute of Experimental Physics, Slovak Academy of Sciences, Ko\v{s}ice, Slovakia
\and
Laboratori Nazionali di Legnaro, INFN, Legnaro, Italy
\and
Secci\'{o}n F\'{\i}sica, Departamento de Ciencias, Pontificia Universidad Cat\'{o}lica del Per\'{u}, Lima, Peru
\and
Lawrence Livermore National Laboratory, Livermore, CA, United States
\and
Division of Experimental High Energy Physics, University of Lund, Lund, Sweden
\and
Universit\'{e} de Lyon 1, CNRS/IN2P3, Institut de Physique Nucl\'{e}aire de Lyon, Lyon, France
\and
Centro de Investigaciones Energ\'{e}ticas Medioambientales y Tecnol\'{o}gicas (CIEMAT), Madrid, Spain
\and
Instituto de Ciencias Nucleares, Universidad Nacional Aut\'{o}noma de M\'{e}xico, Mexico City, Mexico
\and
Instituto de F\'{\i}sica, Universidad Nacional Aut\'{o}noma de M\'{e}xico, Mexico City, Mexico
\and
Centro de Investigaci\'{o}n y de Estudios Avanzados (CINVESTAV), Mexico City and M\'{e}rida, Mexico
\and
Institute for Nuclear Research, Academy of Sciences, Moscow, Russia
\and
Institute for Theoretical and Experimental Physics, Moscow, Russia
\and
Moscow Engineering Physics Institute, Moscow, Russia
\and
Russian Research Centre Kurchatov Institute, Moscow, Russia
\and
Indian Institute of Technology, Mumbai, India
\and
Institut f\"{u}r Kernphysik, Westf\"{a}lische Wilhelms-Universit\"{a}t M\"{u}nster, M\"{u}nster, Germany
\and
SUBATECH, Ecole des Mines de Nantes, Universit\'{e} de Nantes, CNRS-IN2P3, Nantes, France
\and
Yale University, New Haven, CT, United States
\and
Budker Institute for Nuclear Physics, Novosibirsk, Russia
\and
Oak Ridge National Laboratory, Oak Ridge, TN, United States
\and
Physics Department, Creighton University, Omaha, NE, United States
\and
Institut de Physique Nucl\'{e}aire d'Orsay (IPNO), Universit\'{e} Paris-Sud, CNRS-IN2P3, Orsay, France
\and
Department of Physics, University of Oslo, Oslo, Norway
\and
Dipartimento di Fisica dell'Universit\`{a} and Sezione INFN, Padova, Italy
\and
Sezione INFN, Padova, Italy
\and
Faculty of Nuclear Sciences and Physical Engineering, Czech Technical University in Prague, Prague, Czech Republic
\and
Institute of Physics, Academy of Sciences of the Czech Republic, Prague, Czech Republic
\and
Institute for High Energy Physics, Protvino, Russia
\and
Benem\'{e}rita Universidad Aut\'{o}noma de Puebla, Puebla, Mexico
\and
Pusan National University, Pusan, South Korea
\and
Nuclear Physics Institute, Academy of Sciences of the Czech Republic, \v{R}e\v{z} u Prahy, Czech Republic
\and
Dipartimento di Fisica dell'Universit\`{a} `La Sapienza' and Sezione INFN, Rome, Italy
\and
Sezione INFN, Rome, Italy
\and
Commissariat \`{a} l'Energie Atomique, IRFU, Saclay, France
\and
Dipartimento di Fisica `E.R.~Caianiello' dell'Universit\`{a} and Sezione INFN, Salerno, Italy
\and
California Polytechnic State University, San Luis Obispo, CA, United States
\and
Departamento de F\'{\i}sica de Part\'{\i}culas and IGFAE, Universidad de Santiago de Compostela, Santiago de Compostela, Spain
\and
Universidade de S\~{a}o Paulo (USP), S\~{a}o Paulo, Brazil
\and
Russian Federal Nuclear Center (VNIIEF), Sarov, Russia
\and
Department of Physics, Sejong University, Seoul, South Korea
\and
Yonsei University, Seoul, South Korea
\and
Technical University of Split FESB, Split, Croatia
\and
V.~Fock Institute for Physics, St. Petersburg State University, St. Petersburg, Russia
\and
Institut Pluridisciplinaire Hubert Curien (IPHC), Universit\'{e} de Strasbourg, CNRS-IN2P3, Strasbourg, France
\and
University of Tokyo, Tokyo, Japan
\and
Dipartimento di Fisica dell'Universit\`{a} and Sezione INFN, Trieste, Italy
\and
Sezione INFN, Trieste, Italy
\and
University of Tsukuba, Tsukuba, Japan
\and
Dipartimento di Fisica Sperimentale dell'Universit\`{a} and Sezione INFN, Turin, Italy
\and
Sezione INFN, Turin, Italy
\and
Institute for Subatomic Physics, Utrecht University, Utrecht, Netherlands
\and
Soltan Institute for Nuclear Studies, Warsaw, Poland
\and
Warsaw University of Technology, Warsaw, Poland
\and
Purdue University, West Lafayette, IN, United States
\and
Zentrum f\"{u}r Technologietransfer und Telekommunikation (ZTT), Fachhochschule Worms, Worms, Germany
\and
Hua-Zhong Normal University, Wuhan, China
\and
Yerevan Physics Institute, Yerevan, Armenia
\and
Rudjer Bo\v{s}kovi\'{c} Institute, Zagreb, Croatia
}

\date{}

\abstract{The pseudorapidity density and multiplicity distribution of charged particles produced in proton--proton collisions at the LHC, at a centre-of-mass energy $\cms = 7$~TeV, were measured in the central pseudorapidity region \etain{1}. Comparisons are made with previous measurements at $\cms = 0.9$~TeV and 2.36~TeV. At $\cms = 7$~TeV, for events with at least one charged particle in \etain{1}, we obtain $\dNdEta = 6.01 \pm 0.01 (\emph{stat.}) ^{+0.20}_{-0.12} (\emph{syst.})$. This corresponds to an increase of $57.6\,\% \pm 0.4\,\%(\emph{stat.}) ^{+3.6}_{-1.8}\,\%(\emph{syst.})$ relative to collisions at 0.9~TeV, significantly higher than calculations from commonly used models. The multiplicity distribution at 7~TeV is described fairly well by the negative binomial distribution.}

\maketitle
%
%

\section*{Introduction}

We present the pseudorapidity density and the multiplicity distribution for primary charged particles\footnote{Primary particles are defined as prompt particles produced in the collision and all decay products, except products from weak decays of strange particles.} from a sample of $3 \times 10^5$ proton--proton events at a centre-of-mass energy $\cms = 7$~TeV collected with the ALICE detector~\cite{ALICEdet} at the LHC~\cite{LHC}, and compare them with our previous results at $\cms = 0.9$~TeV and $\cms = 2.36$~TeV~\cite{ALICEfirst,ALICEsecond}.
The present study is for the central pseudorapidity region \etain{1}.

In the previous measurements, the main contribution to systematic uncertainties came from the limited knowledge of cross sections and kinematics of diffractive processes. At 7~TeV, there is no experimental information available about these processes; therefore, we do not attempt to normalize our results to the classes of events used in our previous publications (inelastic events and non-single-diffractive events). Instead, we chose an event class  requiring at least one charged particle in the pseudorapidity interval \etain{1} (INEL$>$0$_{|\eta|<1}$), minimizing the model dependence of the corrections. We re-analyzed the data already published at 0.9~TeV and 2.36~TeV in order to normalize the results to this event class. These measurements have been compared to calculations with several commonly used models~\cite{Pythia,Pythia1,D6Ttune,CSCtune,Perugiatune,PhoJet} which will allow a better tuning to accurately simulate minimum-bias and underlying-event effects. Currently, the expectations for 7~TeV differ significantly from one another, both for the average multiplicity and for the multiplicity distribution (see e.g.~\cite{Grosse}).

\section*{ALICE detector and data collection}

The ALICE detector is described in~\cite{ALICEdet}. This analysis uses data from the Silicon Pixel Detector (SPD) and the VZERO counters, as described in~\cite{ALICEfirst,ALICEsecond}. The SPD detector comprises two cylindrical layers (radii 3.9~cm and 7.6~cm) surrounding the central beam pipe, and covers the pseudorapidity ranges $|\eta| < 2$ and $|\eta| < 1.4$, for the inner and outer layers, respectively.
The two VZERO scintillator hodoscopes are placed on either side of the interaction region at $z = 3.3$~m and $z = -0.9$~m, covering the pseudorapidity regions $2.8 < \eta< 5.1$ and $-3.7 < \eta < -1.7$, respectively.

Data were collected at a magnetic field of 0.5~T. The typical bunch intensity for collisions at 7~TeV was\linebreak $1.5 \times 10^{10}$ protons resulting in a luminosity around\linebreak $10^{27}$~cm$^{-2}$s$^{-1}$. There was only one bunch per beam colliding at the ALICE interaction point. 
The probability that a recorded event contains more than one collision was estimated to be around $2 \times 10^{-3}$. A consistent value was measured by counting the events where more than one distinct vertex could be reconstructed.
We checked that pileup events did not introduce a significant bias using a simulation.

 The data at 0.9~TeV and 7~TeV were collected with a trigger requiring a hit in the SPD or in either one of the VZERO counters; i.e. essentially at least one charged particle anywhere in the 8 units of pseudorapidity. At 2.36~TeV, the VZERO detector was turned off; the trigger required at least one hit in the SPD (\etain{2}). The events were in coincidence with signals from two beam pick-up counters, one on each side of the interaction region, indicating the passage of proton bunches. Control triggers taken (with the exception of the 2.36~TeV data) for various combinations of beam and empty-beam buckets
 were used to measure beam-induced and accidental backgrounds. Most backgrounds were removed as described in~\cite{ALICEsecond}.
The remaining background in the sample is of the order of $10^{-4}$ to $10^{-5}$ and can be neglected.

\section*{Event selection and analysis}

The position of the interaction vertex was reconstructed by correlating hits in the two silicon-pixel layers. The vertex resolution achieved depends on the track multiplicity, and is typically 0.1--0.3~mm in the longitudinal ($z$) and 0.2--0.5~mm in the transverse direction.

The analysis is based on using hits in the two SPD layers to form short track segments, called tracklets. A tracklet is defined by a hit combination, one hit in the inner and one in the outer SPD layer, pointing to the reconstructed vertex. The tracklet algorithm is described in~\cite{ALICEfirst,ALICEsecond}.

Events used in the analysis were required to have a reconstructed vertex and at least one SPD tracklet with \etain{1}.
We restrict the $z$-vertex range  to $|z|<5.5$~cm to ensure that the $\eta$-interval is entirely within the SPD acceptance.
After this selection, 47\,000, 35\,000, and 240\,000 events remain for analysis, at 0.9~TeV, 2.36~TeV, and 7~TeV, respectively.
The selection efficiency was studied using two different Monte Carlo event generators, PYTHIA 6.4.21~\cite{Pythia,Pythia1} tune Perugia-0~\cite{Perugiatune} and PHOJET~\cite{PhoJet}, with detector simulation and reconstruction.

The number of primary charged particles is estimated by counting the number of SPD tracklets, corrected for:
\begin{itemize}
\item{geometrical acceptance and detector and reconstruction efficiencies;}
\item{contamination from weak-decay products of strange particles, gamma conversions, and secondary interactions;}
\item{undetected particles below the 50~MeV/$c$ transverse-momentum cut-off, imposed by absorption in the material;}
\item{combinatorial background in tracklet reconstruction.}
\end{itemize}
The total number of collisions corresponding to our data is obtained from the number of events selected for the analysis, applying corrections for trigger and selection efficiencies. This leads to overall corrections of 7.8\,\%, 7.2\,\%, and 5.7\,\% at 0.9~TeV, 2.36~TeV, and 7~TeV, respectively.

The multiplicity distributions were measured for \etain{1} at each energy.
For the 0.9~TeV and 2.36~TeV data we did not repeat the multiplicity-distribution analysis, we use the results from~\cite{ALICEsecond} while removing the zero-multiplicity bin. 
At 7~TeV, we used the same method as described in~\cite{ALICEsecond,JanFiete} to correct the raw measured distributions for efficiency, acceptance, and other detector effects, which is based on unfolding using a detector response matrix from Monte Carlo simulations. The unfolding procedure applies $\chi^2$ minimization with regularization~\cite{blobel_unfolding}.
Consistent results were found when changing the regularization term and the convergence criteria within reasonable limits, and when using a different unfolding method based on Bayes's theorem~\cite{agostini_bayes,agostini_yellowreport}.

\section*{Systematic uncertainties}

Only events with at least one tracklet in $|\eta|<1$ have been selected for analysis in order to reduce sensitivity to model-dependent corrections.
However, a fraction of diffractive reactions also falls into this event category and influences the correction factors at low multiplicities.
In order to evaluate this effect, we varied the fractions of single-diffractive and double-diffractive events produced by the event generators by $\pm 50$\,\% of their nominal values at 7~TeV, and for the other energies we used the variations described in~\cite{ALICEsecond}. The resulting contributions to the systematic uncertainties are estimated to be 0.5\,\%, 0.3\,\%, and 1\,\% for the data at 0.9~TeV, 2.36~TeV, and 7~TeV, respectively. For the same reason, the event selection efficiency is sensitive to the differences between models used to calculate this correction. Therefore, we used the two models which have the largest difference in their multiplicity distributions at very low multiplicities (see below): PYTHIA tune Perugia-0 and PHOJET. The first one was used to calculate the central values for all our results, and the second for asymmetric systematic uncertainties. The values obtained for this contribution are $+0.8$\,\%, $+1.5$\,\%, and $+2.8$\,\% for the three energies considered.

Other sources of systematic uncertainties, e.g. the particle composition, the $p_T$ spectrum and the detector efficiency, are described in~\cite{ALICEsecond}, and their contributions were estimated in the same way.
As a consequence of the smaller uncertainties on the event selection corrections the total systematic uncertainties are significantly smaller than in our previous analyses, which use as normalization inelastic and non-single-diffractive collisions.
Many of the systematic uncertainties cancel when the ratios between the different energies are calculated, in particular the dominating ones, such as the detector efficiency and the event generator dependence. The systematic uncertainty related to diffractive cross sections was assumed to be uncorrelated between energies.

\begin{table*}[htb]
\centering
\caption{Charged-particle pseudorapidity densities at central pseudorapidity ($|\eta|<1$), for inelastic collisions having at least one charged particle in the same region (INEL$>$0$_{|\eta|<1}$), at three centre-of-mass energies. For ALICE, the first uncertainty is statistical and the second is systematic. The relative increases between the 0.9~TeV and 2.36~TeV data, and between the 0.9~TeV and 7~TeV data, are given in percentages. The experimental measurements are compared to the predictions from models. For PYTHIA the tune versions are given in parentheses. The correspondence is as follows: D6T tune (109), ATLAS-CSC tune (306), and Perugia-0  tune (320).}
\label{multab}
\begin{tabular}{cccccc}
  \hline\noalign{\smallskip}
  Energy & ALICE  & \multicolumn{3}{c}{PYTHIA~\cite{Pythia,Pythia1}} & PHOJET~\cite{PhoJet} \\
  \noalign{\smallskip}\cline{3-5}\noalign{\smallskip}
    (TeV)&        &  (109)~\cite{D6Ttune} &   (306)~\cite{CSCtune}   &   (320)~\cite{Perugiatune} &  \\
  \noalign{\smallskip}\hline\noalign{\smallskip}
   & \multicolumn{5}{c}{Charged-particle pseudorapidity density}\\
  \noalign{\smallskip}\hline\noalign{\smallskip}
  0.9    &   $3.81 \pm 0.01 ^{+0.07}_{-0.07}$ & 3.05 & 3.92 & 3.18 & 3.73 \\
  \noalign{\smallskip}
  2.36   &   $4.70 \pm 0.01 ^{+0.11}_{-0.08}$ & 3.58 & 4.61 & 3.72 & 4.31 \\
  \noalign{\smallskip}
  7      &   $6.01 \pm 0.01 ^{+0.20}_{-0.12}$ & 4.37 & 5.78 & 4.55 & 4.98 \\
  \noalign{\smallskip}\hline\noalign{\smallskip}
   & \multicolumn{5}{c}{Relative increase (\%)}\\
  \noalign{\smallskip}\hline\noalign{\smallskip}
  0.9--2.36 & $23.3 \pm 0.4 ^{+1.1}_{-0.7}$ & 17.3 & 17.6 & 17.3 & 15.4 \\
  \noalign{\smallskip}
  0.9--7    & $57.6 \pm 0.4 ^{+3.6}_{-1.8}$ & 43.0 & 47.6 & 43.3 & 33.4 \\
  \noalign{\smallskip}\hline
\end{tabular}
\end{table*}

\begin{figure}[hbt]
\centering
\includegraphics[width=\linewidth]{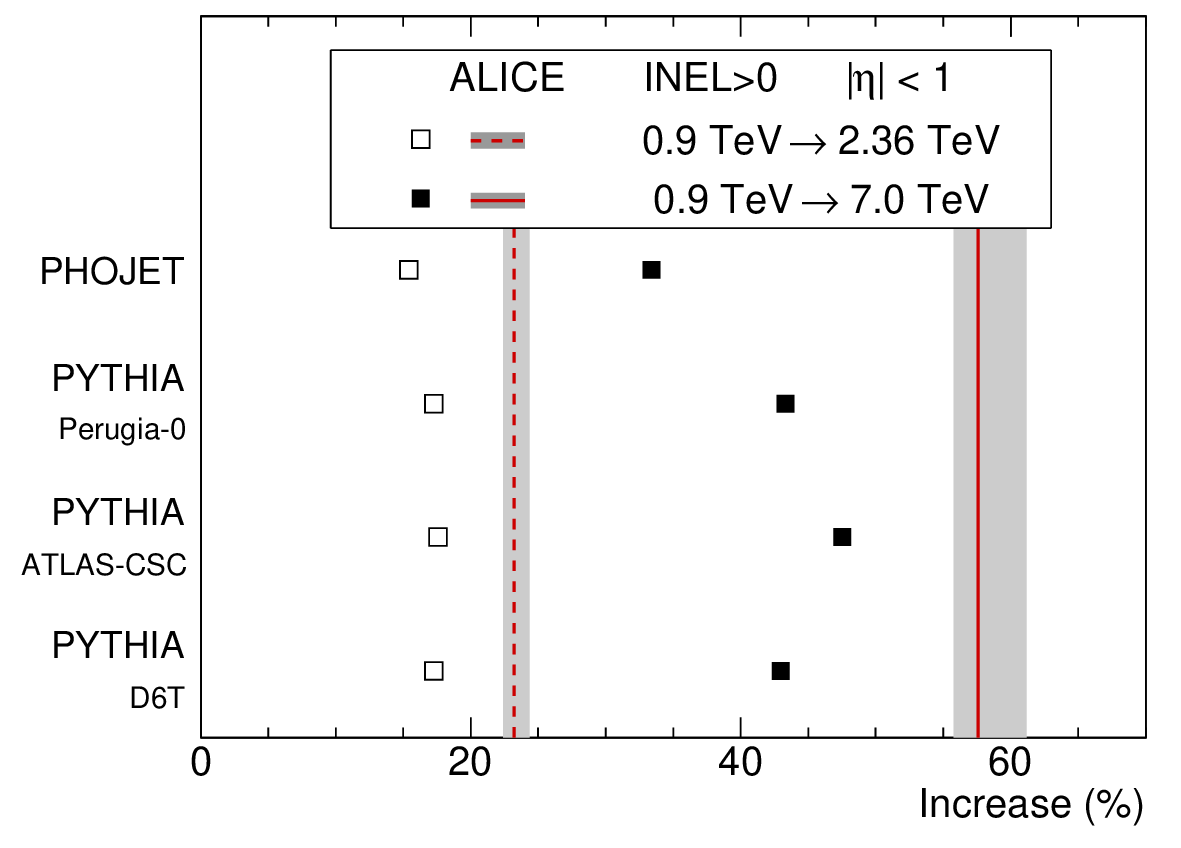}
\caption{Relative increase of the charged-particle pseudorapidity density, for inelastic collisions having at least one charged particle in \etain{1}, between $\cms =0.9$~TeV and 2.36~TeV (open squares) and between $\cms = 0.9$~TeV and 7~TeV (full squares), for various models. Corresponding ALICE measurements are shown with vertical dashed and solid lines; the width of shaded bands correspond to the statistical and systematic uncertainties added in quadrature.}
\label{increasefig}
\end{figure}

\begin{figure}[hbt]
\centering
\includegraphics[width=\linewidth]{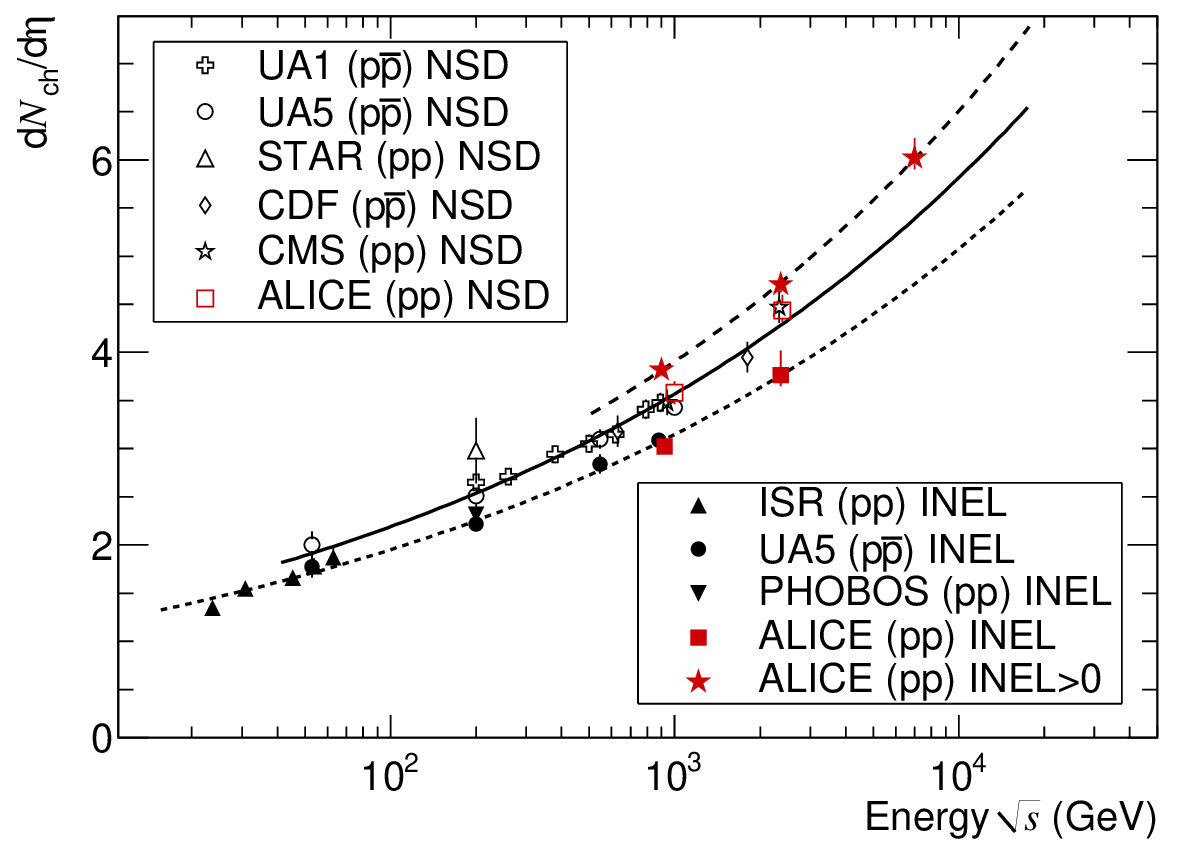}
\caption{Charged-particle pseudorapidity density in the central pseudorapidity region \etain{0.5}
 for inelastic and non-single-diffractive collisions~\cite{ALICEsecond, CMS_first, UA5, ITS_dNdEta, R210,R210p, RHICRef, RHICRef1,UA5Rep, UA1, CDF_dNdEta}, and in \etain{1} for inelastic collisions with at least one charged particle in that region (INEL$>$0$_{|\eta|<1}$), as a function of the centre-of-mass energy. The lines indicate the
fit using a power-law dependence on energy. Note that data points at the same energy have been slightly shifted horizontally for visibility.}
\label{energyfig}
\end{figure}

\section*{Results}

The pseudorapidity density of primary charged particles in the central pseudorapidity region \etain{1} are presented in Table~\ref{multab} and compared to models. The measured values are higher than those from the models considered, except for PYTHIA tune ATLAS-CSC for the 0.9~TeV and 2.36~TeV data, and PHOJET for the 0.9~TeV data, which are consistent with the data. At 7~TeV, the data are significantly higher than the values from the models considered, with the exception of PYTHIA tune ATLAS-CSC, for which the data are only two standard deviations higher. We have also studied the relative increase of pseudorapidity densities of charged particles (Table~\ref{multab}) between the measurement at 0.9~TeV and the measurements at 2.36~TeV and 7~TeV.  We observe an increase of $57.6\,\% \pm 0.4\,\%(\emph{stat.}) ^{+3.6}_{-1.8}\,\%(\emph{syst.})$ between the 0.9~TeV and 7~TeV data, compared with an increase of 47.6\,\% obtained from the closest model, PYTHIA tune ATLAS-CSC (Fig.~\ref{increasefig}). The 7~TeV data confirm the observation made in~\cite{ALICEsecond,CMS_first} that the measured multiplicity density increases with increasing energy significantly faster than in any of the models considered.

\begin{figure*}[tb]
\centering
  \includegraphics[width=\columnwidth]{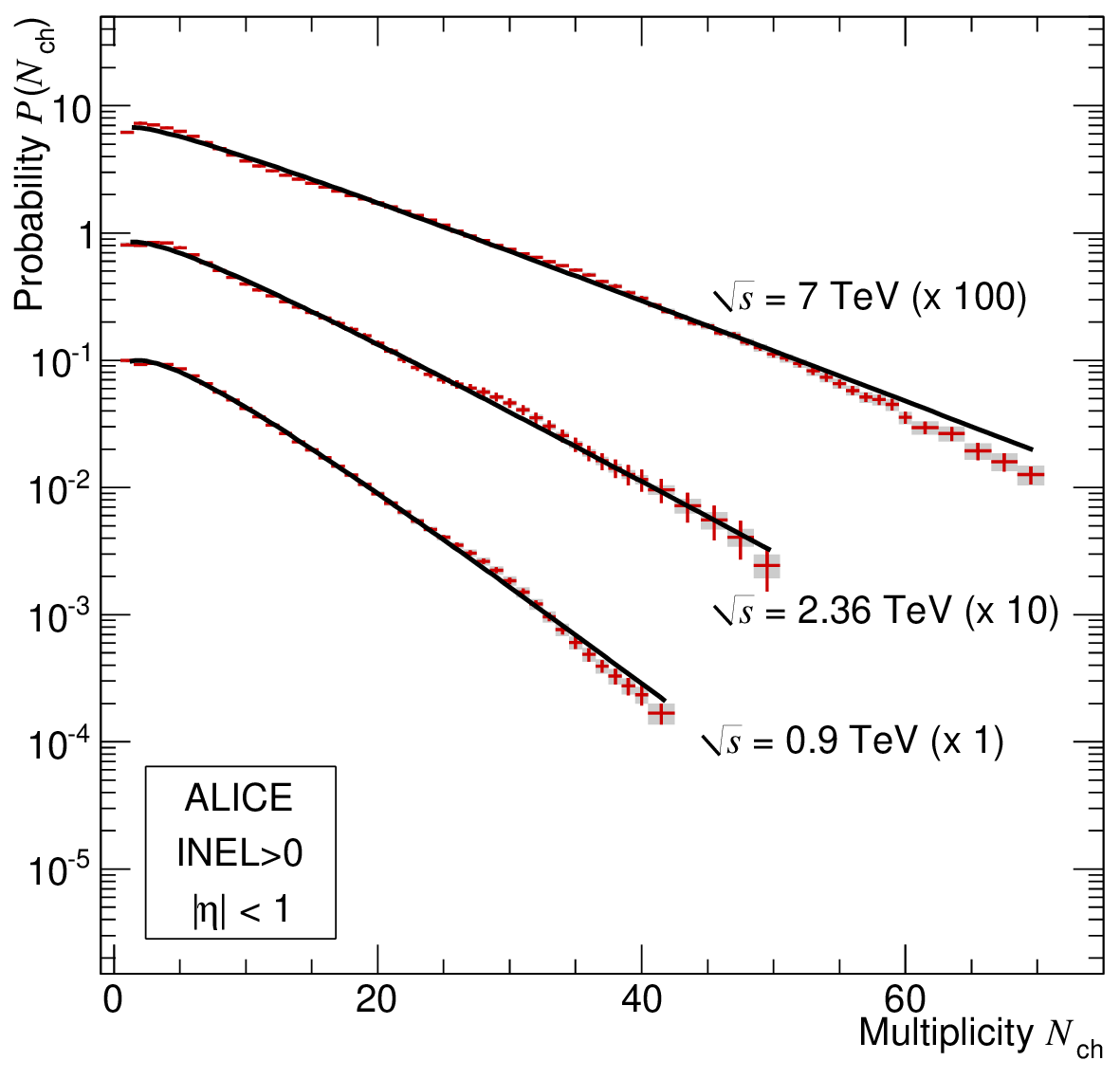}
  \includegraphics[width=\columnwidth]{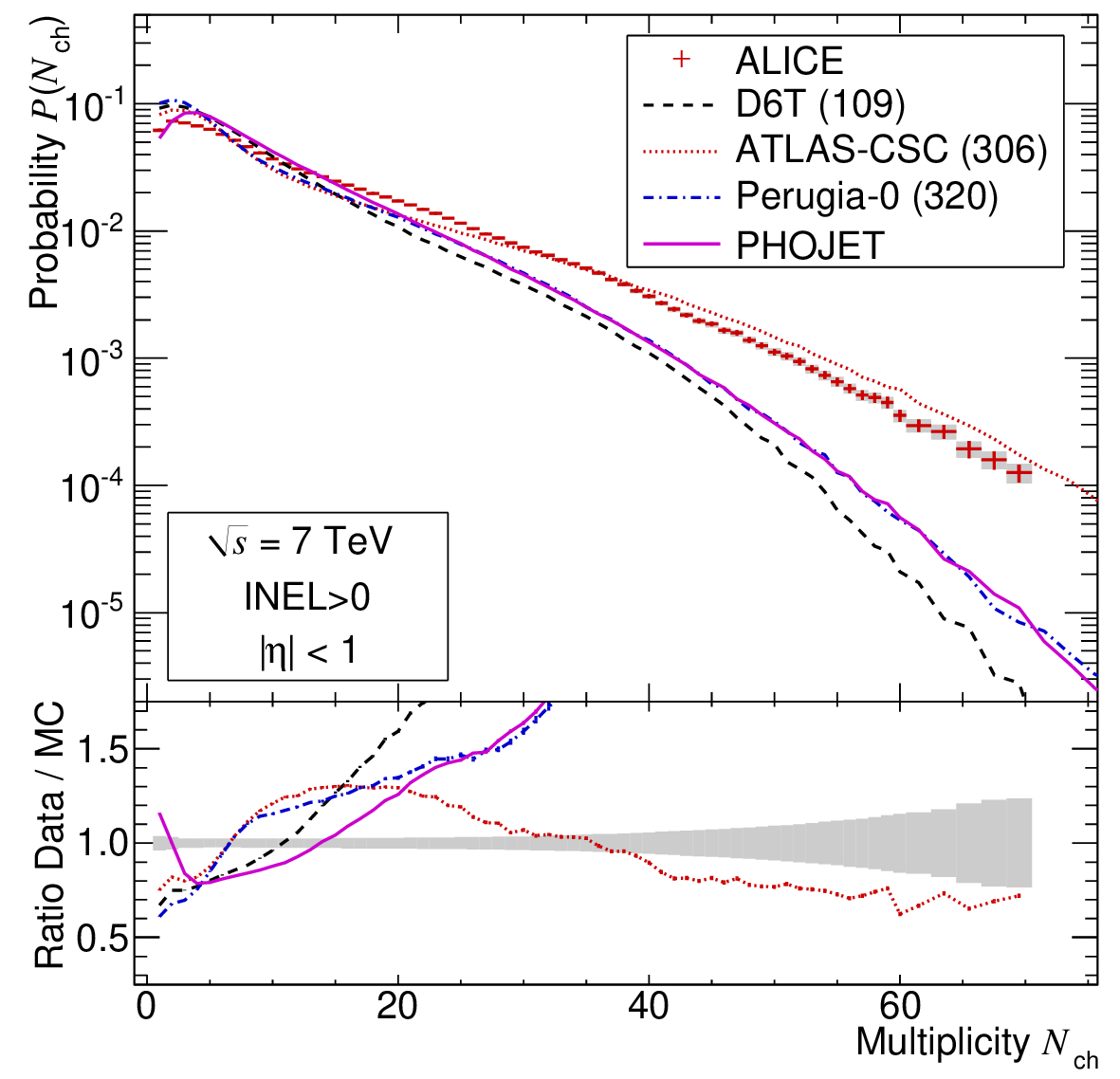}
  \caption{Measured multiplicity distributions in \etain{1} for the INEL$>$0$_{|\eta|<1}$ event class. The error bars for data points represent statistical uncertainties, the shaded areas represent systematic uncertainties. Left: The data at the three energies are shown with the NBD fits (lines). Note that for the 2.36~TeV and 7~TeV data the distributions have been scaled for clarity by the factors indicated. Right: The data at 7~TeV are compared to models: PHOJET (solid line), PYTHIA tunes D6T (dashed line), ATLAS-CSC (dotted line) and Perugia-0 (dash-dotted line). In the lower part, the ratios between the measured values and model calculations are shown with the same convention. The shaded area represents the combined statistical and systematic uncertainties.}
  \label{distfig}
\end{figure*}

 In Fig.~\ref{energyfig}, we compare the centre-of-mass energy dependence of the pseudorapidity density of charged particles for the INEL$>$0$_{|\eta|<1}$ class to the evolution for other event classes (inelastic and non-single-diffractive events), which have been measured at lower energies. Note that INEL$>$0$_{|\eta|<1}$ values are higher than inelastic and  non-single-diffractive values, as expected, because events with no charged particles in \etain{1} are removed.

The increase in multiplicity from 0.9~TeV to 2.36~TeV and 7~TeV was studied by measuring the multiplicity distributions for the event class, INEL$>$0$_{|\eta|<1}$ (Fig.~\ref{distfig} left). 
Small wavy fluctuations are seen at multiplicities above 25. While visually they may appear to be significant, one should note that the errors in the deconvoluted distribution are correlated over a range comparable to the multiplicity resolution and the uncertainty bands should be seen as one-standard-deviation envelopes of the deconvoluted distributions (see also~\cite{ALICEsecond}).
The unfolded distributions at 0.9~TeV and 2.36~TeV are described well by the Negative Binomial Distribution (NBD). At 7~TeV, the NBD fit slightly underestimates the data at low multiplicities ($N_\mathrm{ch} < 5$) and slightly overestimates the data at high multiplicities ($N_\mathrm{ch} > 55$).

A comparison of the 7~TeV data with models (Fig.~\ref{distfig} right) shows that only the PYTHIA tune ATLAS-CSC is close to the data at high multiplicities ($N_\mathrm{ch} > 25$). However, it does not reproduce the data in the intermediate multiplicity region ($8 < N_\mathrm{ch} < 25$). At low multiplicities, ($N_\mathrm{ch} < 5$), there is a large spread of values between different models: PHOJET is the lowest and PYTHIA tune Perugia-0 the highest.

\section*{Conclusion}

We have presented measurements of the pseudorapidity density and multiplicity distributions of primary charged particles produced in proton--proton collisions at the LHC, at a centre-of-mass energy $\cms = 7$~TeV. The measured value of the pseudorapidity density at this energy is significantly higher than that obtained from current models, except for PYTHIA tune ATLAS-CSC. The increase of the pseudorapidity density with increasing centre-of-mass energies is significantly higher than that obtained with any of the used models and tunes.

The shape of our measured multiplicity distribution is not reproduced by any of the event generators considered. The discrepancy does not appear to be concentrated in a single region of the distribution, and varies with the model.

\begin{acknowledgement}

\section*{Acknowledgements}

The ALICE collaboration would like to thank all its engineers and technicians for their invaluable contributions to the construction of the experiment and the CERN accelerator teams for the outstanding performance of the LHC complex.

The ALICE collaboration acknowledges the following funding agencies for their support in building and
running the ALICE detector:
\begin{itemize}
\item{}
Calouste Gulbenkian Foundation from Lisbon and Swiss Fonds Kidagan, Armenia;
\item{}
Conselho Nacional de Desenvolvimento Cient\'{\i}fico e Tecnol\'{o}gico (CNPq), Financiadora de Estudos e Projetos (FI\-N\-EP),
Funda\c{c}\~{a}o de Amparo \`{a} Pesquisa do Estado de S\~{a}o Paulo (FAPESP);
\item{}
National Natural Science Foundation of China (NSFC), the Chinese Ministry of Education (CMOE)
and the Ministry of Science and Technology of China (MSTC);
\item{}
Ministry of Education and Youth of the Czech Republic;
\item{}
Danish Natural Science Research Council, the Carlsberg Foundation and the Danish National Research Foundation;
\item{}
The European Research Council under the European Community's Seventh Framework Programme;
\item{}
Helsinki Institute of Physics and the Academy of Finland;
\item{}
French CNRS-IN2P3, the `Region Pays de Loire', `Region Alsace', `Region Auvergne' and CEA, France;
\item{}
German BMBF and the Helmholtz Association;
\item{}
Hungarian OTKA and National Office for Research and Technology (NKTH);
\item{}
Department of Atomic Energy and Department of Science and Technology of the Government of India;
\item{}
Istituto Nazionale di Fisica Nucleare (INFN) of Italy;
\item{}
MEXT Grant-in-Aid for Specially Promoted Research, Ja\-pan;
\item{}
Joint Institute for Nuclear Research, Dubna;
\item{}
Korea Foundation for International Cooperation of Science and Technology (KICOS);
\item{}
CONACYT, DGAPA, M\'{e}xico, ALFA-EC and the HELEN Program (High-Energy physics Latin-American--European Network);
\item{}
Stichting voor Fundamenteel Onderzoek der Materie (FOM) and the Nederlandse Organisatie voor Wetenschappelijk Onderzoek (NWO), Netherlands;
\item{}
Research Council of Norway (NFR);
\item{}
Polish Ministry of Science and Higher Education;
\item{}
National Authority for Scientific Research - NASR (Autontatea Nationala pentru Cercetare Stiintifica - ANCS);
\item{}
Federal Agency of Science of the Ministry of Education and Science of Russian Federation, International Science and
Technology Center, Russian Academy of Sciences, Russian Federal Agency of Atomic Energy, Russian Federal Agency for Science and Innovations and CERN-INTAS;
\item{}
Ministry of Education of Slovakia;
\item{}
CIEMAT, EELA, Ministerio de Educaci\'{o}n y Ciencia of Spain, Xunta de Galicia (Conseller\'{\i}a de Educaci\'{o}n),
CEA\-DEN, Cubaenerg\'{\i}a, Cuba, and IAEA (International Atomic Energy Agency);
\item{}
Swedish Reseach Council (VR) and Knut $\&$ Alice Wallenberg Foundation (KAW);
\item{}
Ukraine Ministry of Education and Science;
\item{}
United Kingdom Science and Technology Facilities Council (STFC);
\item{}
The United States Department of Energy, the United States National
Science Foundation, the State of Texas, and the State of Ohio.
\end{itemize}
\end{acknowledgement}

%
\end{document}